\documentclass[aps,pra,twocolumn,showpacs]{revtex4}
\usepackage{amsfonts}
\usepackage{graphicx}
\usepackage{latexsym}

\newcommand{\ket}[1]{\left | #1 \right \rangle}
\newcommand{\bra}[1]{\left \langle #1 \right |}

\newcommand{\hilbert}{{\cal H}}
\begin{document}
\title{Maximal violation of Bell inequality for any given two-qubit pure state}

\author{Yang Xiang}
\email{njuxy@sohu.com}
\affiliation{School of Physics and Electronics, Henan University, Kaifeng 475004, Henan province, China }
\date{\today}
\begin{abstract}
In the case of bipartite two qubits systems, we derive the analytical expression of bound of Bell operator
for any given pure state. Our result not only manifest some properties of Bell inequality, for example which
may be violated by any pure entangled state and only be maximally violated for a maximally entangled state,
but also give the explicit values of maximal violation for any pure state. Finally we point out that
for two qubits systems there is no mixed state which can produce maximal violation of Bell inequality.

\end{abstract}

\pacs{03.65.Ud, 03.65.Ta}
\maketitle





The violation of Bell inequality \cite{bell} means that quantum mechanics cannot be regarded as
a local realism theory. A more general version of Bell inequality for two qubits was given by
Clauser, Horne, Shimony, and Holt \cite{chsh} (CHSH inequality). The significance of Bell inequality is that
which can move the philosophical debate between Einstein and Bohr into the lab. Intuitively, one would link
the violation of Bell inequality with quantum entanglement. Indeed, as early as $1991$, Gisin \textit{et al.} \cite{gisin1}
pointed out that any pure entangled states can violate Bell inequality.
The relations between violations of some inequalities and entanglement have been formulated as the entanglement witness \cite{ew}.

Consider two observers, Alice and Bob, in two distant laboratories. They share a pair of qubits (we denote qubit $a$
and qubit $b$) which are interact in the past and usually entangled. The two observers can choose to measure one
of two dichotomous observables: $A$ or $A'$ at qubit $a$ and $B$ or $B'$ at qubit $b$. In this work we only
consider traceless spin observables, which can be written as $A=\bf{a}\cdot\bf{\sigma}$ and analogously for
$A', B, B'$. The CHSH inequality is
\begin{eqnarray}
|\langle \mathbb{B}\rangle_{\rho}|\equiv|\langle A\otimes(B+B')+A'\otimes(B-B')\rangle_{\rho}|\leq 2,
\label{chsh}
\end{eqnarray}
where $\mathbb{B}$ is a so-called Bell operator \cite{belloperator}, and $|\langle \mathbb{B}\rangle_{\rho}|$
is the expected value of $\mathbb{B}$ in state $\rho$. The quantum bound (or say Cirel'son bound) of $\mathbb{B}$ is given by
Cirel'son \cite{tsir,landau}, and which has been written as $\sqrt{4+|\langle[A,A']\otimes[B,B']\rangle_{\rho}|}$.
Landau \cite{landau} has pointed out that for any choice of the observables there exists a state $\rho$ for which
Cirel'son bound can be reached. From Cirel'son bound one can see that the maximal violation is $2\sqrt{2}$
and which can be achieved only in the case that Alice and Bob both choose a pair of anticommuting observables.

One naturally ask a question: for any given state $\rho$, what is the bound of Bell operator? In our previous work in $2008$ \cite{xiang}
we derived the analytical expression of the tight upper bound of $|\langle \mathbb{B}\rangle_{\rho}|$ for any given
pure state $\rho$ under the condition that Alice and Bob both perform lcoal vertical measurements. In this work we completely
solve this problem, e.g., we present the analytical bound of Bell operator for any given
pure state $\rho$ without any restrictions on measurements of Alice and Bob.
We find that this bound is entirely dependent on
the entanglement of the given pure state, so which is an admirable measure of entanglement.
Finally we point out that
for two qubits systems there is no mixed state which can produce maximal violation of Bell inequality.

Using that the observables have unit square one gets
\begin{eqnarray}
\mathbb{B}^{2}=4+[A,A']\otimes[B,B'].
\label{b2}
\end{eqnarray}
If we assume that $A=\bf{a}\cdot\bf{\sigma}$, $A'=\bf{a'}\cdot\bf{\sigma}$, $B=\bf{b}\cdot\bf{\sigma}$, $B'=\bf{b'}\cdot\bf{\sigma}$,
where all $\bf{a}$, $\bf{a'}$, $\bf{b}$ and $\bf{b'}$ are unit vectors.
The above equation can be written as $\mathbb{B}^{2}=4-4(\bf{a}\times \bf{a'})\cdot\bf{\sigma}\otimes(\bf{b}\times \bf{b'})\cdot\bf{\sigma}$.
So without losing generality, an arbitrary measurement scheme of Alice and Bob can be written as
\begin{eqnarray}
\mathbb{B}^{2}=(U^{a}\otimes U^{b})^{\dag}(4-4\sin{x}\cdot\sigma_{y}\otimes\sigma_{y})(U^{a}\otimes U^{b}).
\label{b22}
\end{eqnarray}
Where $x$ is an adjustable parameter which is dependent on the choice of measurements of Alice and Bob,
and without losing generality we can assume $x\in[0,\pi]$. $U^{a(b)}$
is an arbitrary unitary operation on $a(b)$, which can be written as \cite{nielsen}
\begin{eqnarray}
U&\equiv&U(\alpha,\beta,\gamma,\delta)\nonumber\\
&=&e^{-i \alpha}\left(
\begin{array}{c}
e^{i (-\beta/2-\delta/2)}\cos\frac{\gamma}{2}~~~~-e^{i (-\beta/2+\delta/2)}\sin\frac{\gamma}{2}\\
\\
e^{i (+\beta/2-\delta/2)}\sin\frac{\gamma}{2}~~~~~~e^{i
(+\beta/2+\delta/2)}\cos\frac{\gamma}{2}
\end{array}\right),
\nonumber\\
\label{u}
\end{eqnarray}
where $\alpha, \beta, \gamma$ and $\delta$ are real numbers.
From Eq. (\ref{b22}) we can obtain the spectral decomposition of operator $|\mathbb{B}|$
as
\begin{eqnarray}
|\mathbb{B}|&=&2\sqrt{1+\sin{x}}(U^{a}\otimes U^{b})^{\dag}\ket{\eta_{1}}\bra{\eta_{1}}(U^{a}\otimes U^{b})\nonumber\\
&&+2\sqrt{1-\sin{x}}(U^{a}\otimes U^{b})^{\dag}\ket{\eta_{2}}\bra{\eta_{2}}(U^{a}\otimes U^{b})\nonumber\\
&&+2\sqrt{1+\sin{x}}(U^{a}\otimes U^{b})^{\dag}\ket{\eta_{3}}\bra{\eta_{3}}(U^{a}\otimes U^{b})\nonumber\\
&&+2\sqrt{1-\sin{x}}(U^{a}\otimes U^{b})^{\dag}\ket{\eta_{4}}\bra{\eta_{4}}(U^{a}\otimes U^{b}),
\label{absb}
\end{eqnarray}
where
\begin{eqnarray}
\eta_{1}=
\frac{1}{\sqrt{2}}\left(\begin{array}{c}1\\0\\0\\1\end{array}\right),
\eta_{2}=\frac{1}{\sqrt{2}} \left(\begin{array}{c}-1\\0\\0\\1\end{array}\right)
\nonumber\\
\eta_{3}=\frac{1}{\sqrt{2}}
\left(\begin{array}{c}0\\-1\\1\\0\end{array}\right),
\eta_{4}=\frac{1}{\sqrt{2}}\left(\begin{array}{c}0\\
1\\1\\0\end{array}\right).\nonumber\\
\label{eigen}
\end{eqnarray}

For any operator $O$ which has spectral decomposition $O=\sum_{a}O_{a}\ket{a}\bra{a}$, the spectral decomposition
of corresponding operator $|O|$ is $|O|=\sqrt{O^2}=\sum_{a}|O_{a}|\ket{a}\bra{a}$, where $O_{a}$'s are eigenvalues of operator $O$
and $\ket{a}$'s are corresponding eigenvectors. We also notice that the partial trace and full trace of
Bell operator $\mathbb{B}$ both equal to zero, so the spectral decomposition of Bell operator $\mathbb{B}$ can only
be one of the following cases:
\begin{eqnarray}
\mathbb{B}&=&\pm2\sqrt{1+\sin{x}}(U^{a}\otimes U^{b})^{\dag}\ket{\eta_{1}}\bra{\eta_{1}}(U^{a}\otimes U^{b})\nonumber\\
&&\pm2\sqrt{1-\sin{x}}(U^{a}\otimes U^{b})^{\dag}\ket{\eta_{2}}\bra{\eta_{2}}(U^{a}\otimes U^{b})\nonumber\\
&&\mp2\sqrt{1+\sin{x}}(U^{a}\otimes U^{b})^{\dag}\ket{\eta_{3}}\bra{\eta_{3}}(U^{a}\otimes U^{b})\nonumber\\
&&\mp2\sqrt{1-\sin{x}}(U^{a}\otimes U^{b})^{\dag}\ket{\eta_{4}}\bra{\eta_{4}}(U^{a}\otimes U^{b}).\nonumber\\
\label{bell operator}
\end{eqnarray}
In this work we only consider the first case in which eigenvectors $\eta_{1},\eta_{2}$ corresponding to positive eigenvalues.
The other case will give same result.

If one apply a proper product unitary operation on qubits $a$ and $b$, a general pure state $\psi$ always can
be written as
\begin{eqnarray}
\ket{\psi}&=&\cos\left(\frac{\theta}{2}\right)\ket{+z}_{a}\ket{-z}_{b}
+e^{i\chi}\sin\left(\frac{\theta}{2}\right)\ket{-z}_{a}\ket{+z}_{b}\nonumber\\
&=&(0,~~~\cos(\theta/2),~~~e^{i\chi}\sin(\theta/2),~~~0)^{T}.
\label{state}
\end{eqnarray}
where $\ket{\pm z}$ denote the eigenvectors of $\sigma_{z}$, and the subscript specifies the related qubit $a$ or $b$.
The ``angle'' $\theta$ in Eq. (\ref{state}) determines the degree of
entanglement in the state. The angle satisfies $0\leq\theta\leq\pi$,
$\theta=0$ and $\theta=\pi$ correspond to the product states, and
the maximal entanglement occurs at $\theta=\frac{\pi}{2}$.

From Eq. (\ref{eigen}), We find that the Hilbert space of the two qubits, $\hilbert
=C^{2}\otimes C^{2}$, can be divided into two disjoint subspaces:
$\hilbert_{1}$($\{\eta_{1},\eta_{2}\}$) and
$\hilbert_{2}$($\{\eta_{3},\eta_{4}\}$). So in order
to get the maximum of $|\langle\mathbb{B}\rangle_{\rho}|$ it is best to find
some $U^{a}\otimes U^{b}$ which can map $\psi$ into either
$\hilbert_{1}$ or $\hilbert_{2}$, and then choose a proper $x$ which maximizing $|\langle\mathbb{B}\rangle_{\rho}|$.

For the convenience of calculation, we present the explicit expression of $U^{a}\otimes U^{b}$. We use $U^{a}(\alpha,\beta,\gamma,\delta)$($U^{b}(\alpha',\beta',\gamma',\delta')$)
to denote an arbitrary unitary operation on $a$($b$), $U^{a}\otimes U^{b}$ can be expressed
as follows:
\begin{widetext}
\begin{eqnarray}
U^{a}\otimes U^{b}=\left(\begin{array}{c}
e^{i\xi_{11}}\cos(\gamma/2)\cos(\gamma'/2)~~-e^{i\xi_{12}}\cos(\gamma/2)\sin(\gamma'/2)~~
-e^{i\xi_{13}}\sin(\gamma/2)\cos(\gamma'/2)~~e^{i\xi_{14}}\sin(\gamma/2)\sin(\gamma'/2)\\
e^{i\xi_{21}}\cos(\gamma/2)\sin(\gamma'/2)~~~~e^{i\xi_{22}}\cos(\gamma/2)\cos(\gamma'/2)
~~~-e^{i\xi_{23}}\sin(\gamma/2)\sin(\gamma'/2)~-e^{i\xi_{24}}\sin(\gamma/2)\cos(\gamma'/2)\\
e^{i\xi_{31}}\sin(\gamma/2)\cos(\gamma'/2)~~-e^{i\xi_{32}}\sin(\gamma/2)\sin(\gamma'/2)~~
e^{i\xi_{33}}\cos(\gamma/2)\cos(\gamma'/2)~~-e^{i\xi_{34}}\cos(\gamma/2)\sin(\gamma'/2)\\
e^{i\xi_{41}}\sin(\gamma/2)\sin(\gamma'/2)~~~~~e^{i\xi_{42}}\sin(\gamma/2)\cos(\gamma'/2)~~~~
e^{i\xi_{43}}\cos(\gamma/2)\sin(\gamma'/2)~~~~~e^{i\xi_{44}}\cos(\gamma/2)\cos(\gamma'/2)
\end{array}\right),
\nonumber\\
\label{uu}
\end{eqnarray}
\end{widetext}
where all $\xi_{ij}$ are related to tunable parameters $\alpha,
\alpha', \beta, \beta', \delta, \delta'$ from Eq. (\ref{u}). Now we consider mapping
$\psi$ into subspace $\hilbert_{1}$, with this purpose we find that in Eq. (\ref{uu})
we must choose $\gamma$ and $\gamma'$ in such a way that $\cos(\gamma'/2)=\sin(\gamma/2)=0$ or
$\sin(\gamma'/2)=\cos(\gamma/2)=0$, because otherwise $\ket{\psi'}=U^{a}\otimes
U^{b}\ket{\psi}$ will
have component state which is in $\hilbert_{2}$ and this will reduce
$|\bra{\psi'}\mathbb{B}\ket{\psi'}|$. When
$\cos(\gamma'/2)=\sin(\gamma/2)=0$, we can obtain
\begin{eqnarray}
\psi'&=&U^{a}\otimes U^{b}\ket{\psi}\nonumber\\
&=&\left(-e^{i\xi_{12}}\cos(\theta/2)~~~~~0~~~~~0~~~~~
e^{i(\xi_{43}+\chi)}\sin(\theta/2)\right)^{T}.\nonumber\\
\label{state2}
\end{eqnarray}
Then we can obtain $|\bra{\psi'}\mathbb{B}\ket{\psi'}|$
\begin{eqnarray}
&&|\bra{\psi'}\mathbb{B}\ket{\psi'}|\nonumber\\
&=&2\sqrt{1+\sin{x}}\cdot |\langle\eta_{1}|\psi'\rangle|^{2}+2\sqrt{1-\sin{x}}\cdot |\langle\eta_{2}|\psi'\rangle|^{2}\nonumber\\
&=&\sqrt{1+\sin{x}}\cdot
\big|-e^{i\xi_{12}}\cos(\theta/2)+e^{i(\xi_{43}+\chi)}\sin(\theta/2)\big|^2\nonumber\\
&&+\sqrt{1-\sin{x}}\cdot
\big|e^{i\xi_{12}}\cos(\theta/2)+e^{i(\xi_{43}+\chi)}\sin(\theta/2)\big|^2\nonumber\\
&=&\sqrt{1+\sin{x}}+\sqrt{1-\sin{x}}\nonumber\\
&&+\cos{(\xi_{12}-\xi_{43}-\chi)}(\sqrt{1-\sin{x}}-\sqrt{1+\sin{x}})\sin{\theta}.\nonumber\\
\label{expect b}
\end{eqnarray}
Since $\sqrt{1-\sin{x}}-\sqrt{1+\sin{x}}<0$ and $\sin{\theta}>0$, we can take $\cos{(\xi_{12}-\xi_{43}-\chi)}=-1$.
In addition, we notice that $(\sqrt{1+\sin{x}}+\sqrt{1-\sin{x}})^{2}+(\sqrt{1+\sin{x}}-\sqrt{1-\sin{x}})^{2}=4$
and $|\sqrt{1+\sin{x}}+\sqrt{1-\sin{x}}|\leq2$ and $|\sqrt{1+\sin{x}}-\sqrt{1-\sin{x}}|\leq2$, so we can
suppose $\sqrt{1+\sin{x}}+\sqrt{1-\sin{x}}=2\cos{\lambda}$ and $\sqrt{1+\sin{x}}-\sqrt{1-\sin{x}}=2\sin{\lambda}$,
and then $|\bra{\psi'}\mathbb{B}\ket{\psi'}|$ can be written as
\begin{eqnarray}
|\bra{\psi'}\mathbb{B}\ket{\psi'}|=2(\cos{\lambda}+\sin{\lambda}\sin{\theta}).
\end{eqnarray}
So by adjusting parameter $\lambda$ (this equal to adjust $x$), we can get the maximum value of $|\bra{\psi'}\mathbb{B}\ket{\psi'}|$
as
\begin{eqnarray}
|\bra{\psi'}\mathbb{B}\ket{\psi'}|_{max}=2\sqrt{1+\sin^{2}{\theta}}.
\end{eqnarray}

\begin{figure}[t]
\includegraphics[width=0.8\columnwidth,
height=0.40\columnwidth]{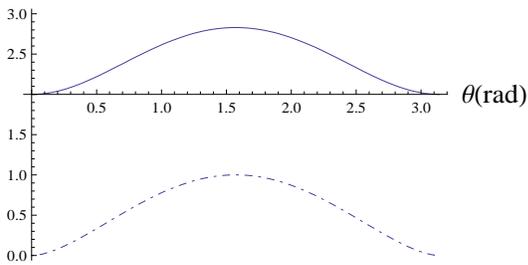} \caption{(Color online). The bound
of $|\langle U^{\dag}\mathbb{B} U\rangle_{\psi}|=2\sqrt{1+\sin^{2}{\theta}}$ (solid line) for any
$\psi$. For any pure entangled state, this bound is greater than the classical bound $2$,
this means that Alice and Bob can achieve the violation of the CHSH inequality. The entanglement $E(\theta)$ (dot-dashed line) of $\psi$ is also
shown. Both the bound and the entanglement are dimensionless
quantities.}
\label{fig1}
\end{figure}

When $\sin(\gamma'/2)=\cos(\gamma/2)=0$, we have
\begin{eqnarray}
\psi'&=&U^{a}\otimes U^{b}\ket{\psi}\nonumber\\
&=&\left(-e^{i(\xi_{13}+\chi)}\sin(\theta/2)~~~~~0~~~~~0~~~~~e^{i\xi_{42}}\cos(\theta/2)
\right)^{T}, \nonumber\\
\label{state3}
\end{eqnarray}
Then we can obtain $|\bra{\psi'}\mathbb{B}\ket{\psi'}|$
\begin{eqnarray}
&&|\bra{\psi'}\mathbb{B}\ket{\psi'}|\nonumber\\
&=&2\sqrt{1+\sin{x}}\cdot |\langle\eta_{1}|\psi'\rangle|^{2}+2\sqrt{1-\sin{x}}\cdot |\langle\eta_{2}|\psi'\rangle|^{2}\nonumber\\
&=&\sqrt{1+\sin{x}}\cdot
\big|-e^{i(\xi_{13}+\chi)}\sin(\theta/2)+e^{i\xi_{42}}\cos(\theta/2)\big|^2\nonumber\\
&&+\sqrt{1-\sin{x}}\cdot
\big|e^{i(\xi_{13}+\chi)}\sin(\theta/2)+e^{i\xi_{42}}\cos(\theta/2)\big|^2\nonumber\\
&=&\sqrt{1+\sin{x}}+\sqrt{1-\sin{x}}\nonumber\\
&&+\cos{(\xi_{13}+\chi-\xi_{42})}(\sqrt{1-\sin{x}}-\sqrt{1+\sin{x}})\sin{\theta}.\nonumber\\
\label{expect b2}
\end{eqnarray}
So we obtain the same maximum of $|\bra{\psi'}\mathbb{B}\ket{\psi'}|$.

We can also choose mapping $\psi$ into subspace $\hilbert_{2}$, and it will produce the same maximum in a similar way.

In Fig. \ref{fig1}, we plot the maximum of $|\bra{\psi'}\mathbb{B}\ket{\psi'}|$, and the entanglement of
$\psi$, which is calculated by using von Neumann entropy of the reduced state in either of the two parties,
\begin{eqnarray}
E(\theta)=-\cos^{2}(\frac{\theta}{2})\log_{2}{\cos^{2}(\frac{\theta}{2})}
-\sin^{2}(\frac{\theta}{2})\log_{2}{\sin^{2}(\frac{\theta}{2})}.\nonumber\\
\end{eqnarray}
We find any entangled pure state can produce violation of Bell inequality, and
only the maximal entangled state can produce the maximal violation of Bell inequality (Cirel'son  bound).

About mixed states, Braunstein \textit{et al.} \cite{braunstein} showed that mixed
states in high dimensional Hilbert space can produce maximal violations of the CHSH inequality, and
the necessary and sufficient condition for violating the CHSH
inequality in an arbitrary mixed spin-$\frac{1}{2}$ state is
presented in \cite{horodecki}. From above calculation about pure states, we find that
in order to produce maximal violations the mixed states of two qubits must be transformable to
any one of $\ket{\eta_{i}}\bra{\eta_{i}}$ in Eq. (\ref{eigen}) by proper unitary operator $U^{a}\otimes U^{b}$,
but all $\ket{\eta_{i}}\bra{\eta_{i}}$ are pure states, so we can conclude that no mixed state of two qubits can
produce maximal violation of the CHSH inequality.

\vskip 0.5 cm

{\it Acknowledgments} This work was supported by National Foundation of Natural Science in
China Grant No. 10947142.


\end{document}